\documentclass[twocolumn]{aastex631}
\usepackage[T1]{fontenc}
\usepackage{amsmath}

\shortauthors{Zheng et al.}

\graphicspath{{./}{figures/}}

\begin{document}

\title{Choked jets in expanding envelope as the origin of the neutrino emission associated with Tidal Disruption Events }

\author[0000-0001-5751-633X]{Jian-He Zheng}
\affiliation{School of Astronomy and Space Science, Nanjing University, Nanjing 210023, People’s Republic of China}
\affiliation{Key laboratory of Modern Astronomy and Astrophysics (Nanjing University), \\
Ministry of Education, Nanjing 210023, People’s Republic of China}

\author[0000-0003-1576-0961]{Ruo-Yu Liu}
\affiliation{School of Astronomy and Space Science, Nanjing University, Nanjing 210023, People’s Republic of China}
\affiliation{Key laboratory of Modern Astronomy and Astrophysics (Nanjing University), \\
Ministry of Education, Nanjing 210023, People’s Republic of China}

\author[0000-0002-5881-335X]{Xiang-Yu Wang}
\affiliation{School of Astronomy and Space Science, Nanjing University, Nanjing 210023, People’s Republic of China}
\affiliation{Key laboratory of Modern Astronomy and Astrophysics (Nanjing University), \\
Ministry of Education, Nanjing 210023, People’s Republic of China}
\email{xywang@nju.edu.cn}

\begin{abstract}

Three tidal disruption event (TDE) candidates (AT2019dsg, AT2019fdr, AT2019aalc) have been
found to be coincident with high-energy astrophysical neutrinos in multi-messenger follow-ups. Recent studies suggest the presence of a quasi-spherical, optically thick envelope around the supermassive black holes in TDEs, resulted from stellar debris after the disruption. The envelope may expand outwardly with a velocity of $\sim 10^4\, {\rm km/s}$, as indicated by the emission line widths. We study whether  the neutrino signal can be explained by  choked relativistic jets inside the expanding envelope. While powerful jets, such as that in Swift J1644+57, can successfully break out from the envelope, those with relatively weak power could be choked by the envelope. Choked jets  can still
accelerate cosmic rays and  produce high-energy neutrinos via interaction with the  thermal photons in the envelope.  We explore the parameter space of the jets that can produce detectable neutrino flux while being choked in the expanding envelope.  
We find that  the cumulative neutrino numbers of AT2019fdr and AT2019aalc are consistent with the expected range imposed by observations, while the allowed parameter space for AT2019dsg is small.
The neutrino time delay relative to the  optical peak time of TDEs can be explained as the jet propagation time in the envelope before being choked. The discovery of TDE-associated neutrino events may suggest that jets might have been commonly formed  in  TDEs, as expected from super-Eddington accretion, but most of them are too weak to break out from the expanding envelopes. 

\end{abstract}


\section{Introduction} \label{sec:intro}
The origin of extragalactic high-energy neutrinos is one of the main puzzles in neutrino astronomy.
Neutrino alert-triggered follow-up searches in electromagnetic data have proven successful in identifying individual  blazars as sources, with the most prominent case being TXS 0506+056,
which was found to be in a gamma-ray flaring state during the neutrino emission \citep{2018ICECUBESci...361.1378I}. In the time-integrated point source searches,  a clustering of neutrinos
from the direction of the starburst galaxy NGC 1068 was found \citep{Icecube2020PhRvL.124e1103A}. Recently, optical follow-up observations of neutrino alerts using the Zwicky
Transient Facility (ZTF) have identified two optical flares from the centers of galaxies coincident with 100 TeV-scale neutrinos: AT2019dsg associated with the IceCube neutrino event IC191001A \citep{2019dsg2021NatAs...5..510S} and AT2019fdr associated with IC200530A \citep{2019fdr}. The former belongs
to the class of spectroscopically-classified tidal disruption events (TDEs) from quiescent black holes, while the latter originates from an unobscured active galactic nucleus (AGN).  Afterward, it was noticed that these TDEs were accompanied by an IR echo, and this connection then led
to the identification of a third TDE, AT2019aalc, as the counterpart of  IC191119A \citep{Velzen2021arXiv211109391V}.
There have been arguments that multiple source populations may contribute to the astrophysical diffuse neutrino flux, based on neutrino event detections and population statistics \citep{Bartos2021ApJ...921...45B}, as well as from spectral shape and directional information \citep{Palladino2018A&A...615A.168P}.

TDEs are phenomena in which a star passes close enough to a supermassive black hole (SMBH) to be ripped apart by its tidal forces.
Relativistic jets  in TDEs have been proposed as ultra-high energy cosmic ray sources \citep{Farrar2009ApJ...693..329F,Farrar2014arXiv1411.0704F}. Neutrino emission is predicted to be produced in both successful jets \citep{Wang2011PhRvD..84h1301W,wang2016PhRvD..93h3005W,Dai2017MNRAS.469.1354D,Lunardini2017PhRvD..95l3001L,Senno2017ApJ...838....3S} and choked jets of TDEs \citep{wang2016PhRvD..93h3005W,Senno2017ApJ...838....3S}.  After the identification of AT2019dsg/IC191001A, TDE jets \citep{Winter2021NatAs...5..472W,Liu2020PhRvD.102h3028L}, corona and hidden wind \citep{Murase2020ApJ...902..108M} as well as  outflow-cloud interactions \citep{Wu2022MNRAS.514.4406W} have been proposed to explain the neutrino emission. While a successful jet has the advantage that it can provide the necessary power for the neutrino emission (see discussion in \cite{Winter2022icrc.confE.997W}), no convincing direct jet signatures for AT2019dsg have been observed \citep{Mohan2022ApJ...927...74M}. For AT2019fdr, similarly, various plausible cosmic ray acceleration sites have been proposed, such as the corona, a subrelativistic wind, or a relativistic jet\citep{2019fdr}.
The neutrino production site is therefore uncertain, and \cite{Winter2023ApJ...948...42W} recently discussed unified  time-dependent interpretations of these events, considering three models in which quasi-isotropic neutrino emission is due to the interactions of accelerated protons of moderate, medium, and high-energy with X-rays, OUV, and IR photons, respectively.

Recent studies suggest the presence of an extended, quasi-spherical, optically thick envelope around the SMBH in TDEs \citep{Loeb1997ApJ...489..573L,Coughlin2014ApJ...781...82C,Roth2016ApJ...827....3R,Metzger2022ApJ...937L..12M}. 
The presence of the gas envelope can solve the puzzle that the temperatures (a few $10^4$ K) found in
optically discovered TDEs are significantly lower than the predicted thermal temperature ($>10^5$K) of the accretion
disk \citep{Gezari2012Natur.485..217G,Wevers2019MNRAS.488.4816W}. The gas at large radii can absorb UV photons produced by the inner accretion disk and re-emit photons at lower temperatures. The TDE flare spectra generally show modest emission line widths of $\sim 10^4 {\rm km/s}$, which may represent  the outflow velocity of the reprocessing envelope \citep{Roth2018ApJ...855...54R}.  

As no convincing  jet signatures for the three TDEs have been observed, 
in this paper, we study whether the three neutrino events can be interpreted in the scenario of choked  jets in TDEs.
Numerical simulations by \cite{DeColle2012ApJ...760..103D} show that  powerful jets can successfully cross the envelope. However, jets without enough power could be choked in the dense envelope  \citep{wang2016PhRvD..93h3005W,DeColle2020NewAR..8901538D}. Three jetted TDEs, Swift J1644+57, Swift J2058+05, and Swift J1112-8238 all have relativistic jets with a non-thermal X-ray luminosity above $10^{47}{\rm erg s^{-1}}$. Since the black hole accretion is expected to produce a continuous distribution of  jet luminosity, as has been seen in AGN jets, the fact that no less powerful TDE jets have been detected so far is a puzzle. One possibility is that jets with luminosity much less than $10^{47}{\rm erg s^{-1}}$ have been choked in the envelopes.

\section{Model description} \label{sec:model}
\subsection{Envelope formation}

In the process of tidal disruptions, nearly half of the stellar mass of the disrupted star is ejected from the SMBH while the other half remains gravitationally bound. The remaining debris interacts with itself  and releases the orbital energy through dissipative process such as shocks, turbulence and convection. Outflows can be driven during the dissipation process \citep[e.g.][]{Rees1988Natur.333..523R,Evans1989ApJ...346L..13E,Coughlin2014ApJ...781...82C,Ryu12020ApJ...904...98R}.

\citet{Loeb1997ApJ...489..573L} suggested that, close to the pericenter, a radiation dominated torus may be formed. 
Farther away from the pericenter, rotation is not important and the intense radiation pressure could disperse the marginally bound gas into a quasi-spherical structure \citep{Loeb1997ApJ...489..573L}. A quasi-spherical emission surface is  supported by recent spectropolarimetry observations of some TDEs \citep[e.g.,][]{Patra2022MNRAS.515..138P}. 
It is  possible  that the envelope is pushed out impulsively by a super-Eddington eruption in the disk\citep{Loeb1997ApJ...489..573L}. The unbound envelope could be accelerated by the extra radiative force at the photosphere due to atomic lines \citep{Loeb1997ApJ...489..573L}.

In addition, an envelope can be formed by the disk wind in super-Eddington accretion.  When the radiative loss of a super-Eddington disk is inefficient, photons are trapped within the disk and the thermal energy released by the accretion is stored in the gas. Heating gas will puff up the disk in the vertical direction. A fraction of debris with lower gravitational binding energy is expelled by the disk wind, forming a quasi-spherical envelope around the SMBH \citep{Strubbe2009MNRAS.400.2070S,Coughlin2014ApJ...781...82C,Miller2015ApJ...805...83M,Metzger2016MNRAS.461..948M}.\citet{Metzger2016MNRAS.461..948M} argued that only a small fraction of stellar mass is accreted to the inner disk closed to the black hole ($f_{\rm in}\ll 1$), while the majority instead becoming unbound in an outflow of velocity $\sim 10^4 {\rm km/s}$. Velocities of $\sim 10^4 {\rm km/s}$ is supported by spectral observations of some TDEs \citep[e.g.,][]{Gezari2012Natur.485..217G,Leloudas2019ApJ...887..218L}.

\begin{figure}
    \centering
    \includegraphics[width=0.45\textwidth]{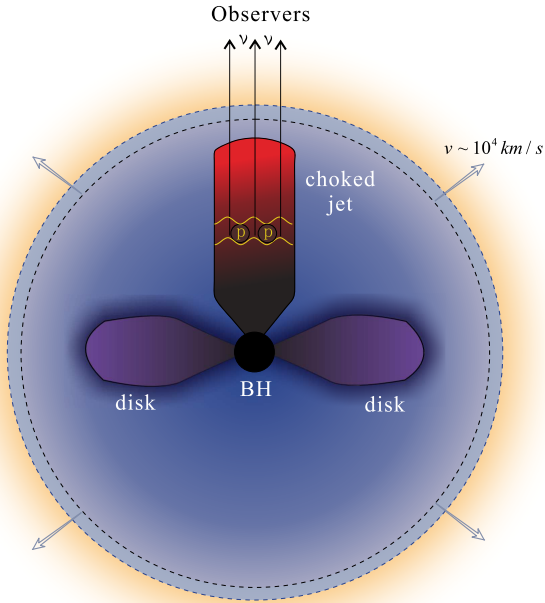}
    \caption{The schematic picture of a choked TDE jet  in an expanding envelope. A quasi-spherical  envelope surrounding the black hole is expanding outwardly with a velocity of $\sim 10^4{\rm km/s}$. Jet is collimated by the cocoon pressure to a cylinder shape and it is choked inside the envelope.  Internal shocks (yellow wavy lines) occurred within the jet accelerate cosmic-ray protons, which produce neutrinos observed by us.}
    \label{fig1}
\end{figure}

We assume that the envelope is spherically symmetric and expanding outwardly. For the structure of an expanding envelope, we follow the work of \cite{Gottlieb2022MNRAS.tmp.2496G}, which simulated the jet propagation in an expanding homologous medium. It extends the existing solution of jets in a static medium to an expanding medium.  The density profile of a homologous expanding envelope is given by \citep{Gottlieb2022MNRAS.tmp.2496G}
\begin{equation}
\rho_{\rm env}(r,t)=K_{\rm env}r^{-n} t^{n-3};\quad v_{\rm min}t<r<v_{\rm max}t,
\end{equation}
where $K_{\rm env}$ is the normalization factor, $t$ is the time since the disruption, $v_{\rm min}$ and $ v_{\rm max}$ are the minimum velocity and maximum velocity of the envelope, respectively. A power index $n<3$ indicates most of mass accumulates in outer layers. For $n=3$, the density at a given radius ($v_{\rm min}t<r<v_{\rm max}t$) is time-invariant, but the profile is still different from static medium since the minimum radius $v_{\rm min}t$ and maximum radius $v_{\rm max}t$ are changing with time.

An exponent $n = 3$ applies to a radiation pressure-supported envelope \citep{Loeb1997ApJ...489..573L}, and $n = 2$ applies to a steady wind outflow \citep{Roth2016ApJ...827....3R}.

\subsection{The condition of choked jets}

As the jet advances in the surrounding envelope, the jet
drives a bow shock ahead of it. The jet is capped by a termination shock, and a reverse shock propagates back into the jet, where the jet is decelerated and heated. 
The velocity of the jet head is determined by the balance of pressure between ambient gas and the jet head \citep{Matzner2003MNRAS.345..575M}. In a Newtonian expanding medium, we can express the velocity in the frame of ejecta by a coordinate transformation \citep[e.g.,][]{Ioka2018PTEP.2018d3E02I,Hamidani2020MNRAS.491.3192H,Gottlieb2022MNRAS.tmp.2496G}:

\begin{equation}
\frac{v_{\rm h}-v_{\rm env,h}}{c}\simeq \frac{1}{1+\tilde{L}^{-1/2}}, 
\end{equation}
where 
\begin{equation}
  \tilde{L}=  \frac{L_{\rm j}}{\Sigma_{\rm j} \rho_{\rm env}(r_{\rm h})c^3} ,
\end{equation}
$L_{\rm j}$ is the luminosity of the jet, $v_{\rm h}$ is the head velocity, $v_{\rm env,h}$ is the envelope velocity at the location of the head (both are measured in the lab frame), and $\Sigma_{\rm j}=\pi r^{2}_{\rm h} \theta^{2}_{\rm h}$ is the jet cross-section at the location of the head with $\theta_{\rm h}$ being the jet head opening angle. Note that the jet may be collimated by the cocoon pressure and $\theta_{\rm h}\le\theta_{j,0}$, where $\theta_{j,0}$ is the initial half open angle of the jet.

For a solar mass envelope, the parameter $\tilde{L}$ is very small $\tilde{L}=0.0025L_{\rm j,45}r_{\rm h,15}M^{-1}_{\rm env,\odot}\theta^{-2}_{\rm h,-1}$, indicating that the jet head is Newtonian $\beta_{\rm h}-\beta_{\rm env,h}\approx\tilde{L}^{1/2}=0.05L^{1/2}_{\rm j,45}r^{1/2}_{\rm h,15}M^{-1/2}_{\rm env,\odot}\theta^{-1}_{\rm h,-1}$.
\cite{Gottlieb2022MNRAS.tmp.2496G} find that in the Newtonian case, the dimensionless jet head velocity $\eta=(v_{\rm h}-v_{\rm env,h})/v_{\rm env,h}$ approaches the value $\eta_{\rm a}=1/(5-n)$ determined by the density profile after enough time. The situation of $\eta\approx\eta_{\rm a}$ is defined as the asymptotic phase of the jet. During the asymptotic phase, the jet head location is independent of initial conditions and can be expressed analytically:

\begin{equation}
{r_{\rm h}} = {\left( {\frac{{2{N_{\rm col}(n)}{{\left[ {2\pi \left( {5 - n} \right)} \right]}^{1/3}}}}{{15}}} \right)^{\frac{3}{{5 - n}}}}{\left( {\frac{{{L_{\rm j}}v_{\max }^{5 - n}{t^{6 - n}}}}{{{E_{\rm ej,kin}}\theta _{\rm j,0}^4}}} \right)^{\frac{1}{{5 - n}}}},
\end{equation}
where $N_{\rm col}(n)$ is a coefficient determined by numerical simulations \citep{Gottlieb2022MNRAS.tmp.2496G}, and $E_{\rm ej,kin}$ is the kinetic energy of the envelope.  

 However, the transition time of the asymptotic phase $t_{\rm a}$ is uncertain, especially for weak jets ($\eta\ll1$). Calculating the transition time $t_{\rm a}$ invokes many detailed parameters of the jet and the ejecta, which is beyond the scope of our work. If we only focus on the breakout criterion rather than the evolution of jet heads, a simple analytic criterion of Newtonian jet heads is proposed by \citet{Gottlieb2022MNRAS.tmp.2496G}
\begin{eqnarray}
    E_{\rm j,iso} &>& \tilde{E}_{\rm a}(n)\frac{(5-n)^{4-n}}{N^{5}_{\rm E}(n)} E_{\rm ej,kin}\theta_{\rm j,0}^2,\\
    &>& \left\{
    \begin{array}{ll}
         299 E_{\rm ej,kin}\theta_{\rm j,0}^2 (n=2)\\
         698 E_{\rm ej,kin}\theta_{\rm j,0}^2 (n=3)
    \end{array}
    \right. ,
\end{eqnarray}
where $\tilde{E}_{\rm a}(n)$ and $N_{\rm E}(n)$ are numerical factors calibrated by simulations, and $E_{\rm j,iso}=\int L_{\rm j,iso}dt$ is the total isotropic equivalent energy of jets.

In the free-fall situation, matter returns to the region near the pericenter radius at a rate $\dot M\propto(t/\tau)^{-5/3}$, where $\tau$ is the characteristic timescale for
initiation of this power-law accretion rate, which is roughly the orbital period of the most bound debris, i.e.,
\begin{equation}
    \tau=\frac{\pi}{M_{\star}}\sqrt{\eta^{2}_{\star}\frac{M_{\rm BH}R^{3}_{\star}}{2G}}=41M^{1/2}_{\rm BH,6}\left(\frac{M_{\star}}{M_\odot}\right)^{-0.1}{\rm day},
\end{equation}
where $\eta_{\star}$ is a correction factor depending on the structure of stars, which is close to  $\eta_{\star}\simeq1$ for solar-mass stars \citep{Phinney1989IAUS..136..543P,TDEreview2021ARA&A..59...21G}.  We adopt the mass-radius relation of stars $R\propto M^{0.6}_{\star}$ from \cite{kippenhahn2012sse..book.....K}.  As the jet power may scale with the accretion rate as $L_{\rm j}\propto \dot M$ in the super-Eddington accretion phase \citep{Krolik2012ApJ...749...92K,Piran2015MNRAS.453..157P}, the characteristic lifetime of the jet  is also $\tau$ and $L_{\rm j}\propto (t/\tau)^{-5/3}$ after that. Noting that the observed power indices $\alpha$ in TDEs are diverse \citep{Velzen2021ApJ9084V},  we take the power index as a free parameter assuming the jet luminosity follows $L_{\rm j}=L_{\rm j,0} (1+t/\tau)^{-\alpha}$, where $L_{\rm j,0}$ is the initial jet luminosity.

From Eq.5, the luminosity of  a choked jet should be
\begin{equation}
    {L_{\rm j,iso,0}} \le \frac{{\left( {\alpha - 1} \right){E_{\rm ej,kin}}\theta_{\rm j,0}^2}}{{\tau \left[ {1 - {{\left( {1 + t_{\rm d}/{\tau }} \right)}^{1 - \alpha}}} \right]}} \times \left\{
    \begin{array}{ll}
         299  (n=2)\\
         698  (n=3)
    \end{array}
    \right.,
\end{equation}
where $t_{\rm d}$ is the neutrino trigger time since the optical peak of TDEs in the observer's frame. We assumed the jet engine works until the neutrino trigger time.

\begin{figure*}
\gridline{\fig{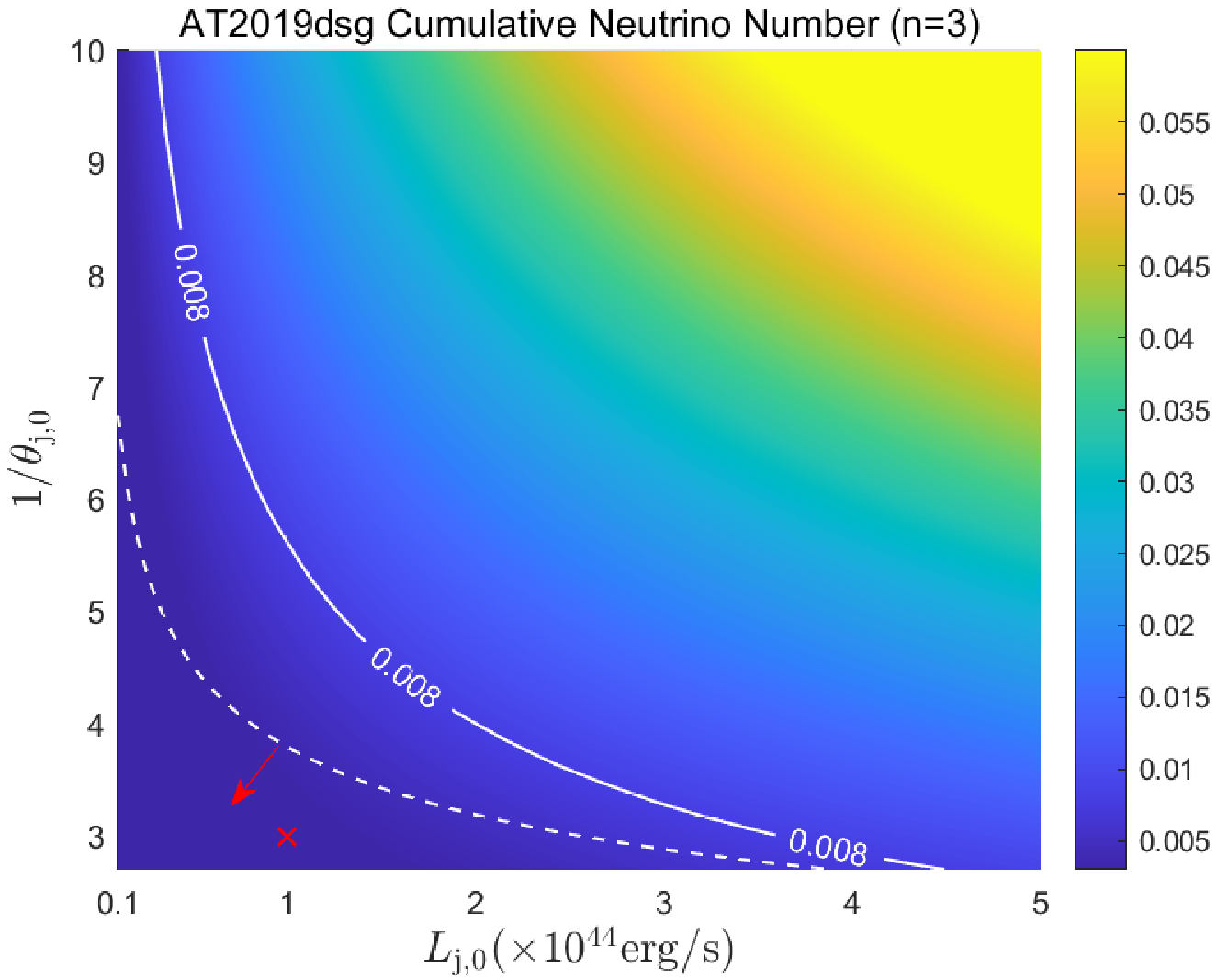}{0.33\textwidth}{(a)}
          \fig{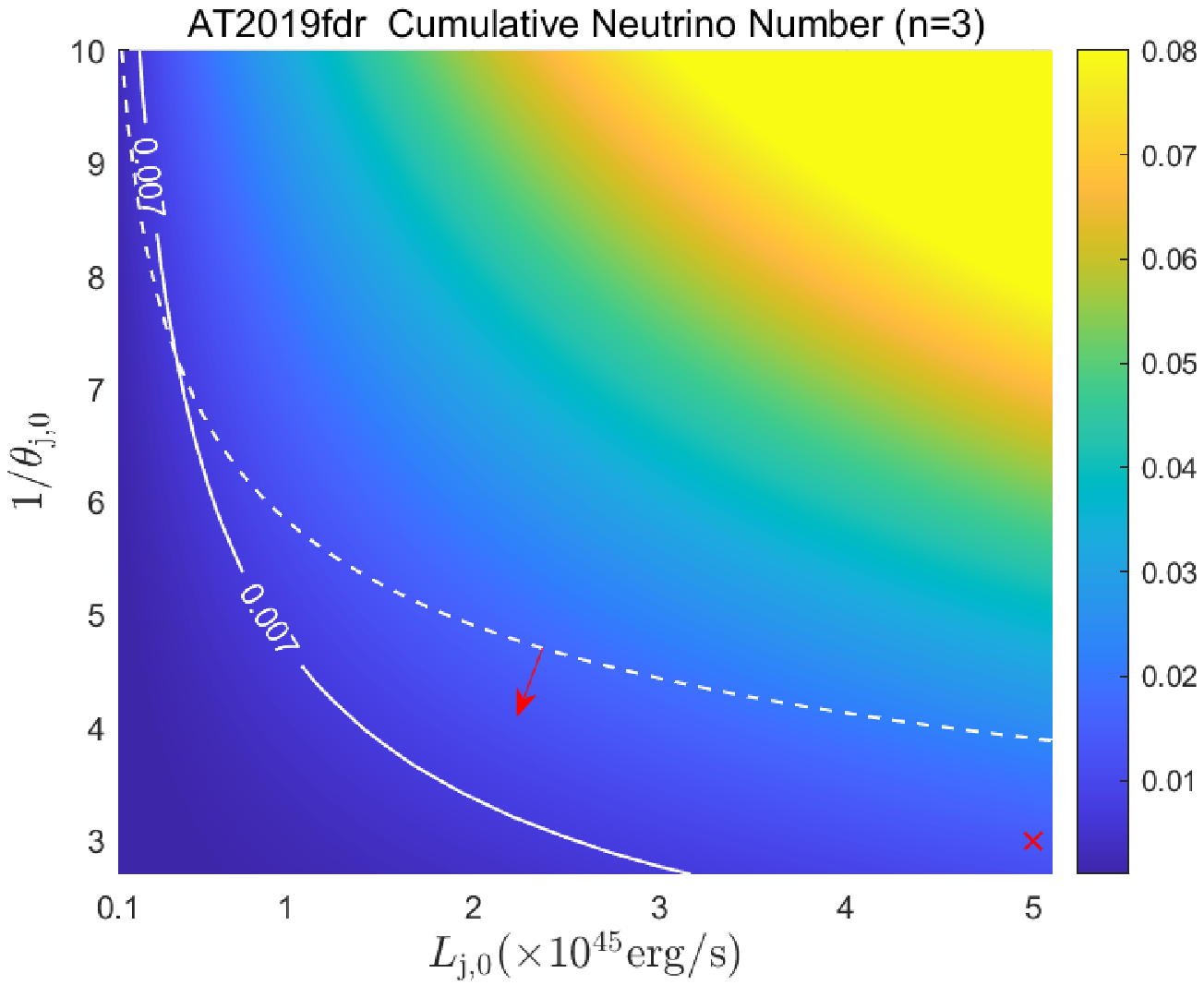}{0.33\textwidth}{(b)}
          \fig{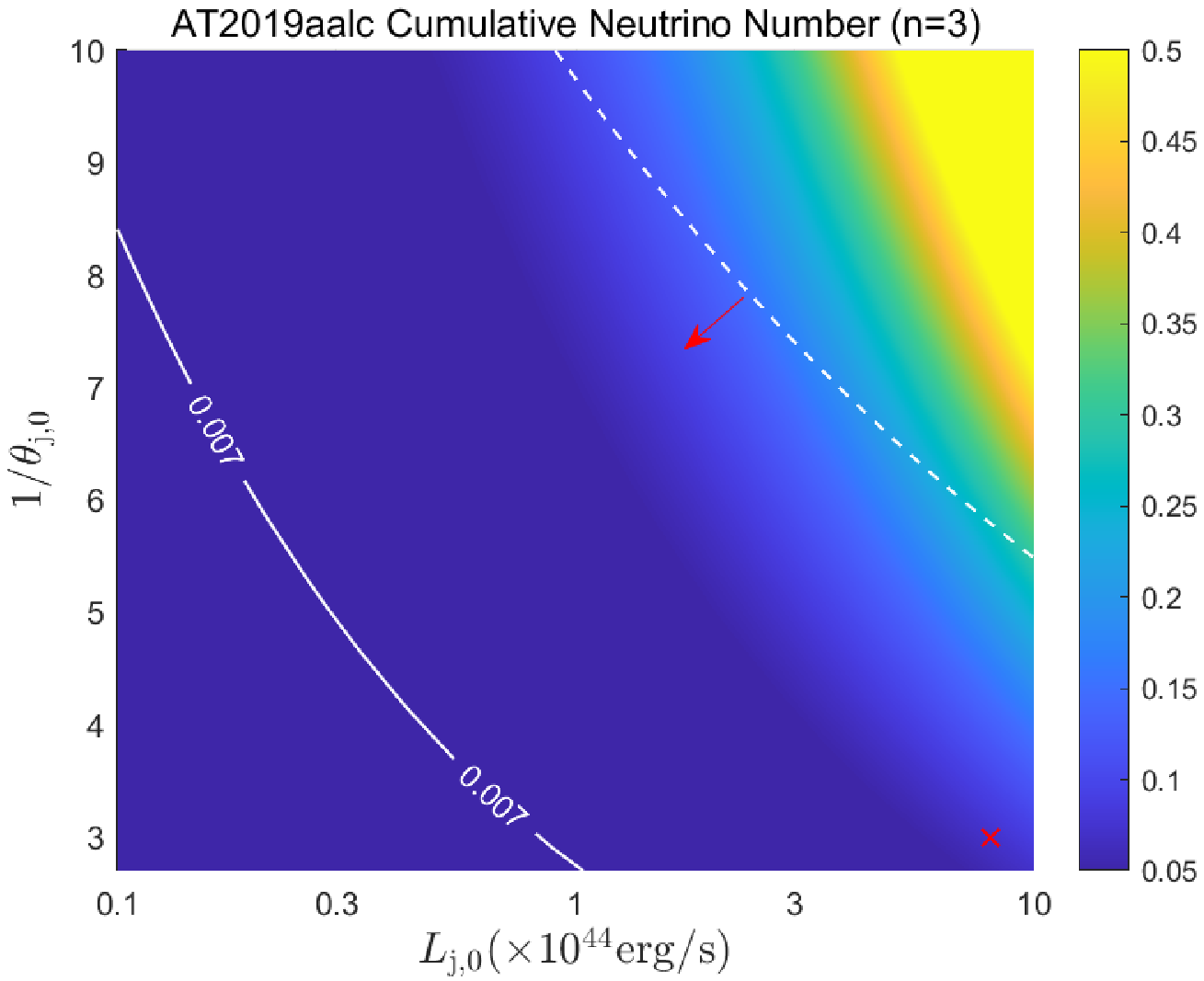}{0.33\textwidth}{(c)}
          }
\gridline{\fig{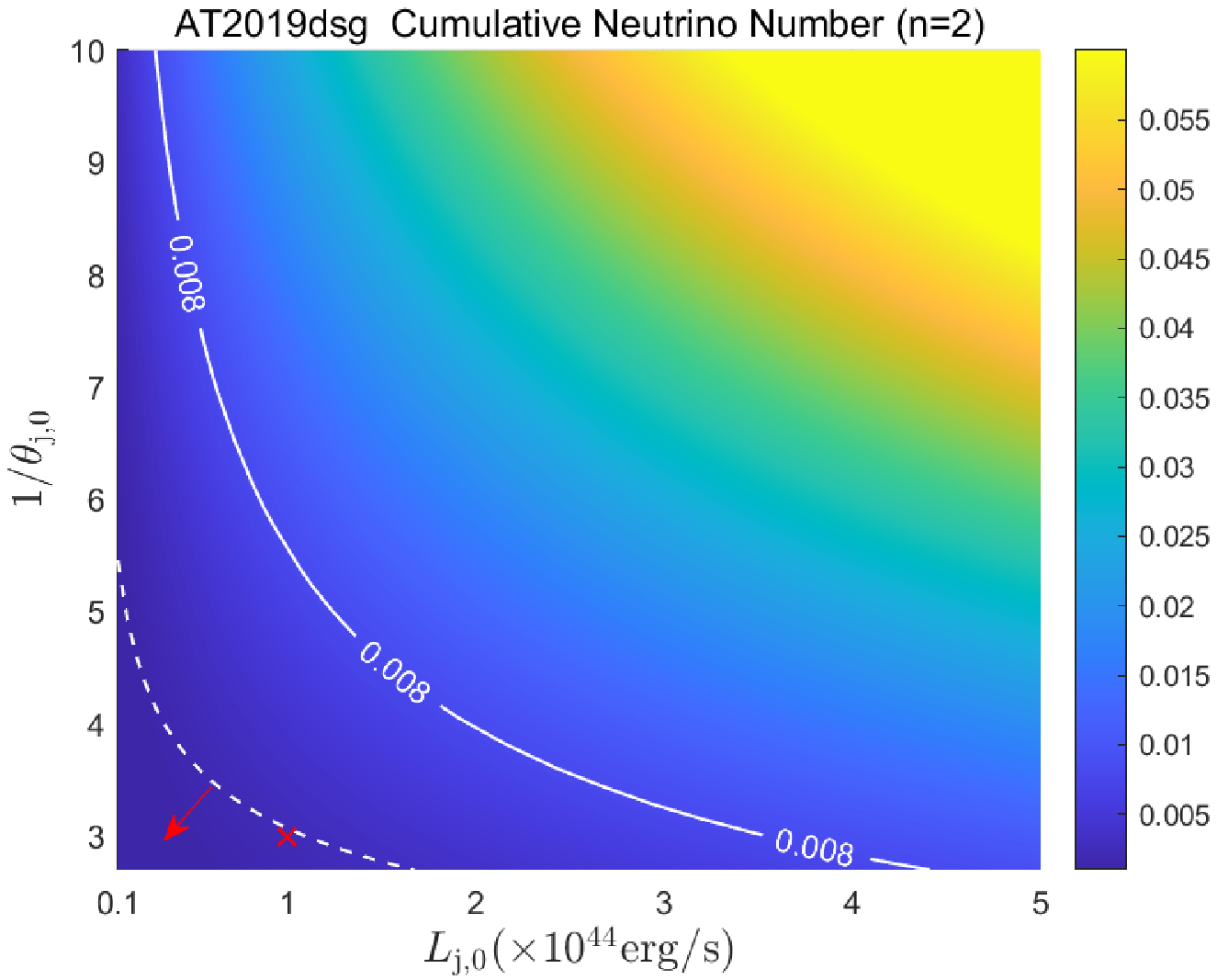}{0.33\textwidth}{(d)}
          \fig{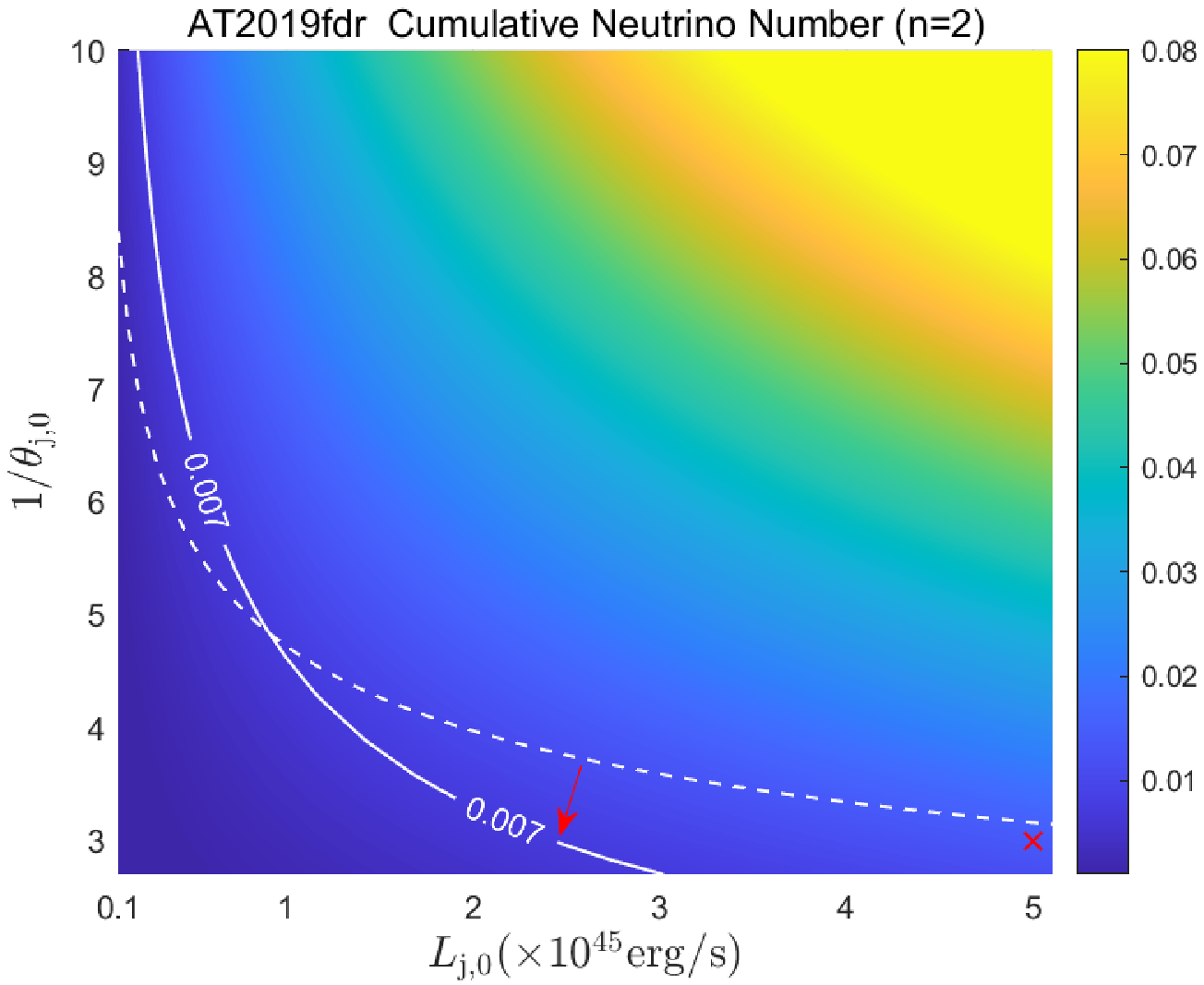}{0.33\textwidth}{(e)}
          \fig{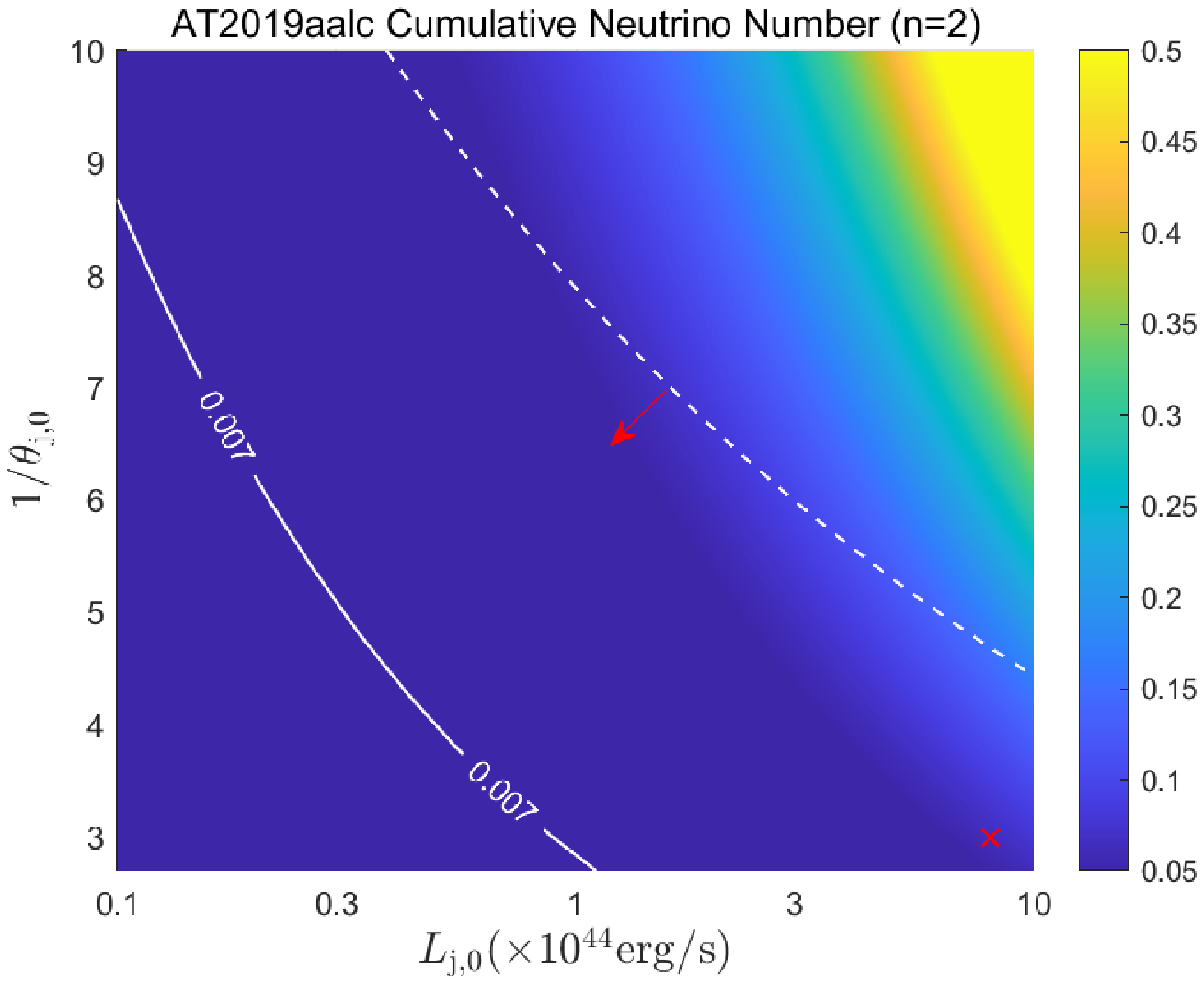}{0.33\textwidth}{(f)}
          }

\caption{The accumulative neutrino numbers (the color map) as a function of the jet luminosity and bulk Lorentz factor assuming envelope masses of three TDEs given in case A. The white solid lines  represent the parameter values corresponding to  the minimally required neutrino numbers, which are $N_{\nu}=0.008$ for AT2019dsg  and $N_{\nu}=0.007$ for AT2019fdr and AT2019aalc \citep{2019dsg2021NatAs...5..510S,2019fdr}. The white dashed line represents the critical parameter values for jets being choked and the direction of the arrow means that the permitted parameter space for choked jets is below this line. Red crosses mark the points in the parameter space that were chosen to produce the results  in Figure \ref{fig:flu}\,\&\ref{fig:num}. The top and bottom panels represent the cases of $n=3$ and $n=2$ for the envelope density profile, respectively. Internal shock radius are taken to be $R_{\rm int}=3\times10^{14}{\rm cm}$ for AT2019aalc and AT2019fdr,  and $R_{\rm int}=10^{14}{\rm cm}$ for AT2019dsg.  $\alpha=5/3,\beta=5/12$, $\theta_{\rm j,0}=1/3$ and $f_{\rm ph}=0.05$ are used in the calculation. Other parameters are listed in Table \ref{parameters}.}
\label{fig:paratest}
\end{figure*}

The dashed lines in Figure \ref{fig:paratest} show the critical values of the jet luminosity, $L_{\rm j,0}=L_{\rm j,iso,0}\theta_{\rm j,0}^2/2$, and the initial opening angle $\theta_{\rm j,0}$ for choked jets in three TDEs. \citet{Bromberg2011ApJ...740..100B} have shown that, in static media, the Lorentz factor above the first collimation shock is $\Gamma=1/\theta_{\rm j,0}$, arising from the requirement that after the collimation the pressure in the shocked jet is similar to the cocoon pressure. The same equality holds also in expanding media.
The typical value of the Lorentz factors is expected to be $\Gamma\sim 10$, as inferred from jetted TDEs  \citep{Metzger2012MNRAS.420.3528M}, so we use a range of $0.1-0.3$ rad for $\theta_{\rm j,0}$.

\begin{table}
  \centering
 \caption{Notations in this paper}
 \begin{tabular}{ccc}
 \hline
 \hline
  Notation & Definition \\
 \hline

 $L_{\rm j}$ & jet luminosity    \\
 $L_{\rm j,iso}$ & jet isotropic luminosity (2$L_{\rm j}/\theta^2_{\rm j,0}$)    \\
 $L_{\rm p}$    &  isotropic proton luminosity ($\varepsilon_{\rm p}L_{\rm j,iso}$)            \\
 $\varepsilon_{\rm p}$&  proton equipartition factor (0.2)            \\
 $\theta_{\rm j,0}$&  initial jet half-opening angle             \\
 $\Gamma$ & jet bulk Lorentz factor ($1/\theta_{\rm j,0}$) \\
 $R_{\rm T}$ & disruption radius      \\
 $R_{\rm int}$ & internal shock radius      \\
 

 \hline
 \end{tabular}
 \label{notation}
\end{table}

\subsection{Neutrino production}
Internal shocks may occur due to the internal collisions within the jets, resulted from the inhomogeneity in the jet velocity.
Temporal variability with $\delta t\approx100s$ has been seen in the X-ray emission of the jetted TDE Swift J1644+57, which is thought to be generated by internal shocks \citep{swiftJ162011Natur.476..421B}. The initial Lorentz factor of the jet is $\Gamma\sim3-10$, and hence the collision occurs at $R_{\rm int}\approx\Gamma^{2}c\delta t=3\times10^{14}{\rm cm}\Gamma_1^2$. 
We assume the jet is matter-dominant when internal shock occurs. The density of jets $n_{\rm j}=1.8\times10^{7}{\rm cm^{-3}}L_{\rm j,iso,46}R^{-2}_{\rm int,14}\Gamma^{-2}_{1}$ is much lower than that of envelopes $n=10^{13}{\rm cm^{-3}}M_{\rm env,\odot}R^{-3}_{\rm int,14} $. Thus, the low-density jet forms a cavity inside the envelope, while the jets are propagating. The envelope contains dense thermal photons, which diffuse out of
the optically thick envelope and enter the optically-thin cavity. The radiation diffusion time in the envelope is $t_{\rm diff}=r^2\kappa \rho(r)/c$ at a particular radius $r$ \citep{Roth2016ApJ...827....3R}. Using $\kappa R_{\rm ph} \rho_{\rm ph}=1$, the diffusion time is  $t_{\rm diff}=R_{\rm ph}/c$ in the case of $n=2$ and $t_{\rm diff}=(R_{\rm ph}/c)(R_{\rm ph}/r)$ in the case of $n=3$. Compared to the jet propagation time ($r/v_{\rm h}$), the number density of photons in the cavity should be suppressed by a factor of $f_{\rm ph}= \min \{ (\frac{c}{v_{\rm h}})(\frac{r}{R_{\rm ph}}), 1\}$ in the case of $n=2$ and  $f_{\rm ph}= \min \{ (\frac{c}{v_{\rm h}})(\frac{r}{R_{\rm ph}})^2, 1\}$ in the case of $n=3$. 
Internal shocks that propagate into the low-density jets are collisionless, although they locate inside the optically thick envelope \citep{wang2016PhRvD..93h3005W}. It has been shown that
shocks in TDE jets can accelerate cosmic rays to ultrahigh energies \citep{Farrar2009ApJ...693..329F,Wang2011PhRvD..84h1301W}.

The accelerated protons will produce neutrinos by $pp$ and $p\gamma$ reactions. The dense envelope and high temperature provide a thick target for neutrino production. The cooling time of $pp$ interaction is $t^{-1}_{pp}=\kappa_{pp }n_{\rm p}\sigma_{pp}c$, where $\kappa_{pp}\approx0.5$ is the inelasticity of protons and $\sigma_{pp}\approx3\times10^{26}{\rm cm^2}$ is the cross section.  Meanwhile, the cooling time of $p\gamma$ interaction is
\begin{equation}
    t^{-1}_{p\gamma}=\frac{c}{2\gamma^2_{\rm p}}\int^{\infty}_{\overline{\epsilon}_{\rm th}}\sigma_{\rm p\gamma}(\overline{\epsilon})\kappa_{\rm p\gamma}(\overline{\epsilon})\overline{\epsilon}d\overline{\epsilon}\int^{\infty}_{\overline{\epsilon}/2\gamma_{\rm p}}\epsilon^{-2}\frac{dn}{d\epsilon}d\epsilon,
\end{equation}
where $\gamma_{\rm p}$ is Lorentz factors of protons, $dn/d\epsilon$ is the number density of seed photons, $\epsilon_{\rm th}\simeq145{\rm MeV}$ is the threshold energy of $p\gamma$ interaction, $\epsilon$ and $\overline{\epsilon}$ are photons' energy in the observer's frame and center of mass frame of protons, respectively. Seed photons of $p\gamma$ interaction are mainly dominated by thermal photons of the envelope. 

\begin{figure*}
\gridline{\fig{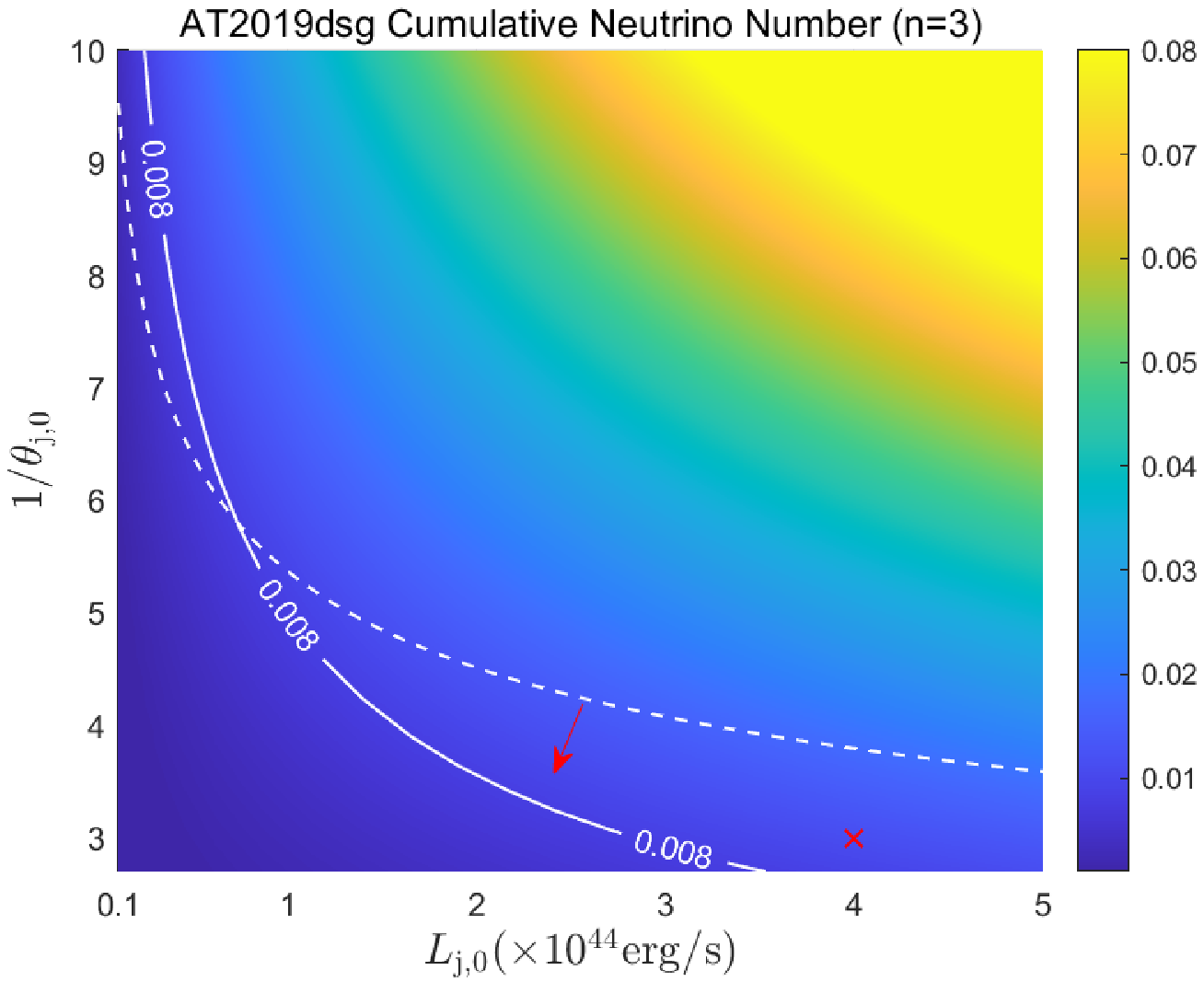}{0.33\textwidth}{(a)}
          \fig{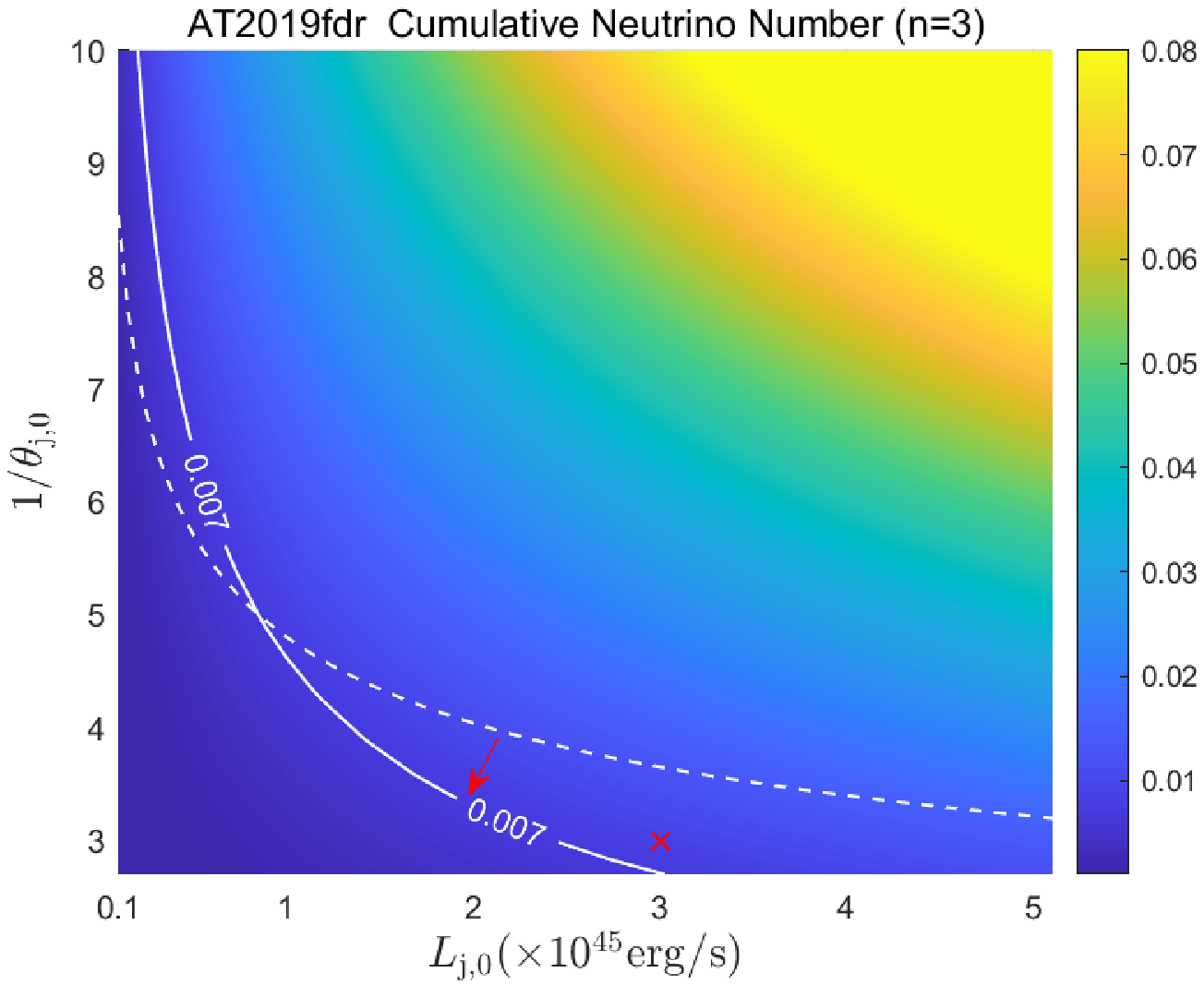}{0.33\textwidth}{(b)}
          \fig{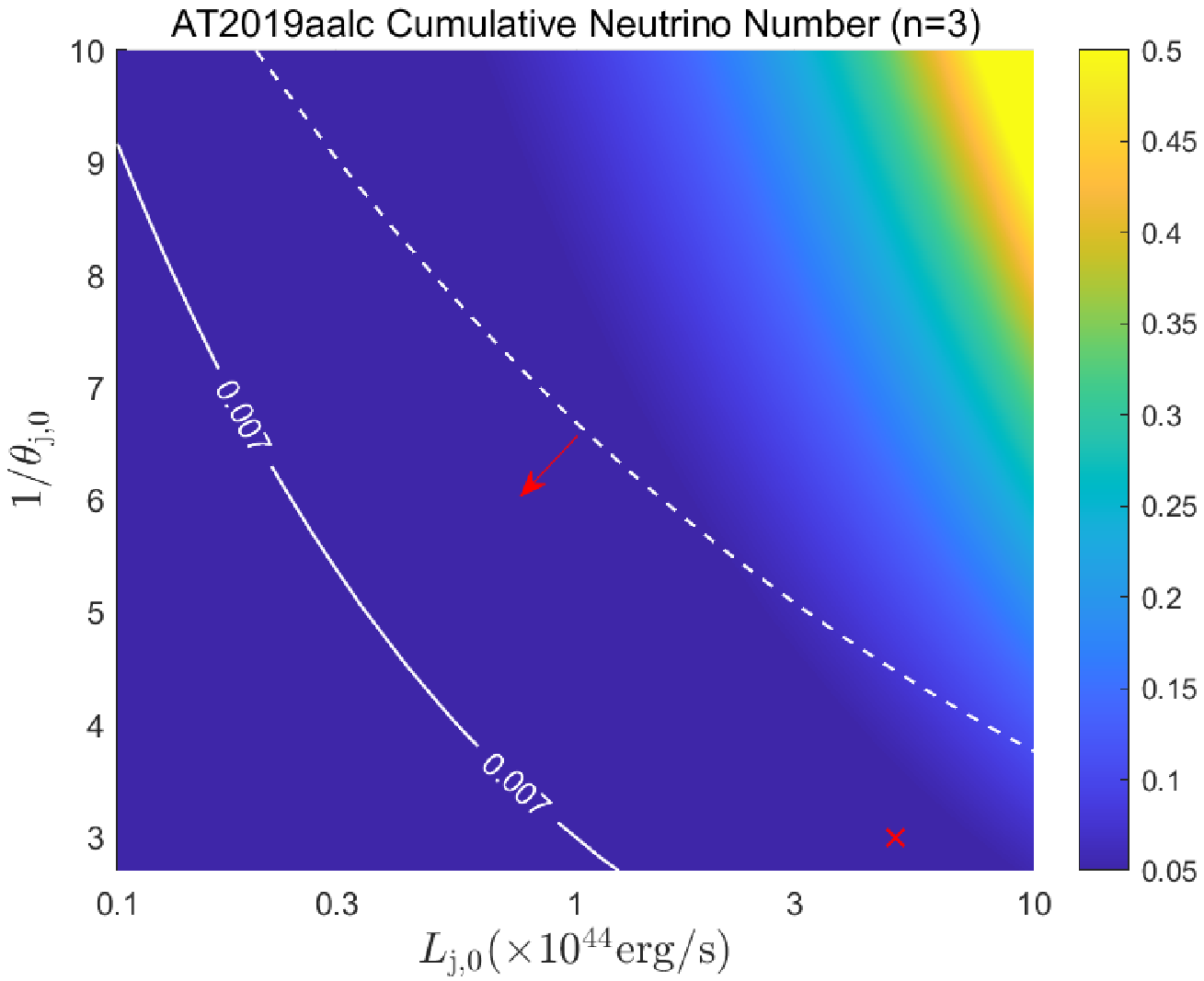}{0.33\textwidth}{(c)}
          }
\gridline{\fig{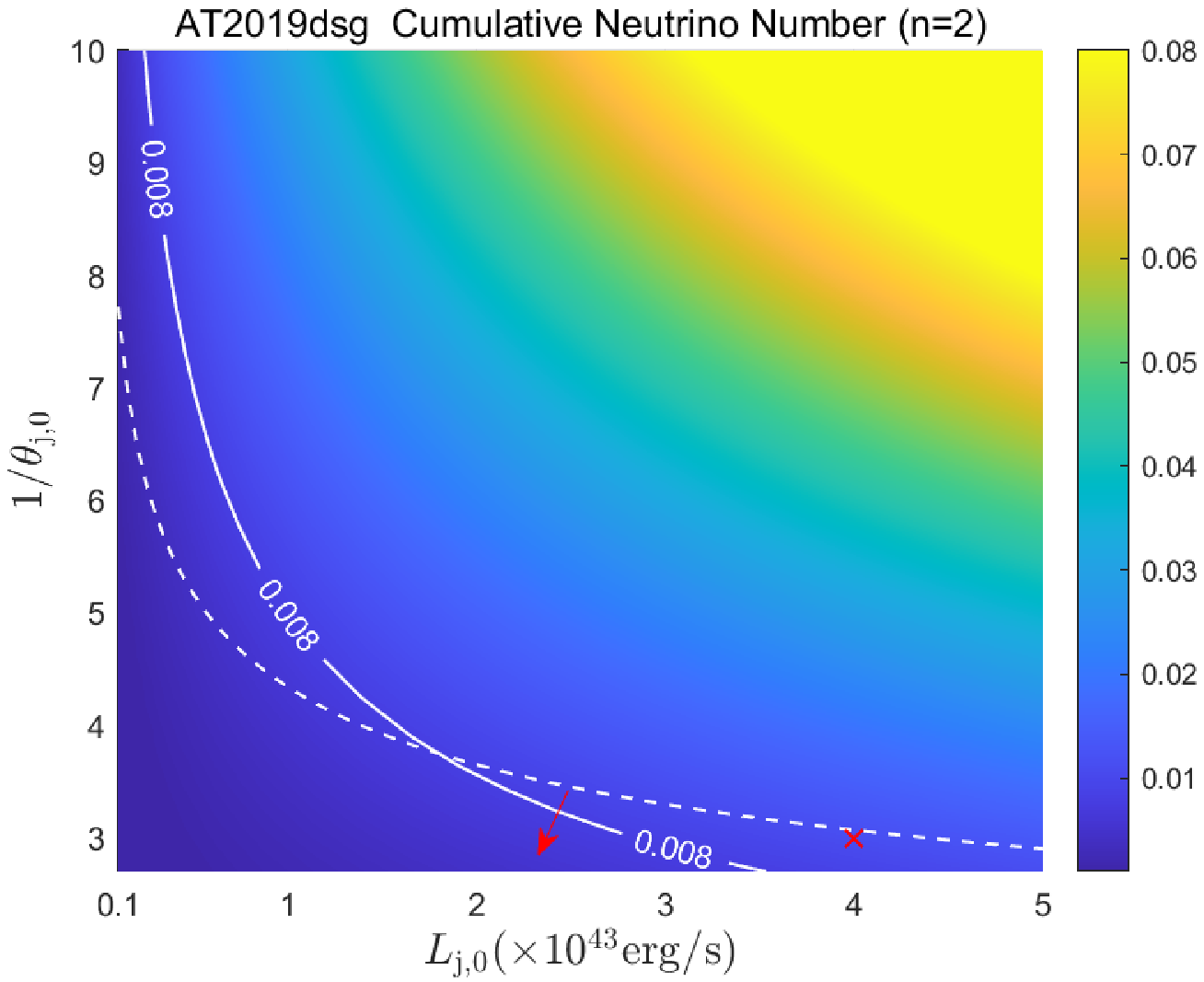}{0.33\textwidth}{(d)}
          \fig{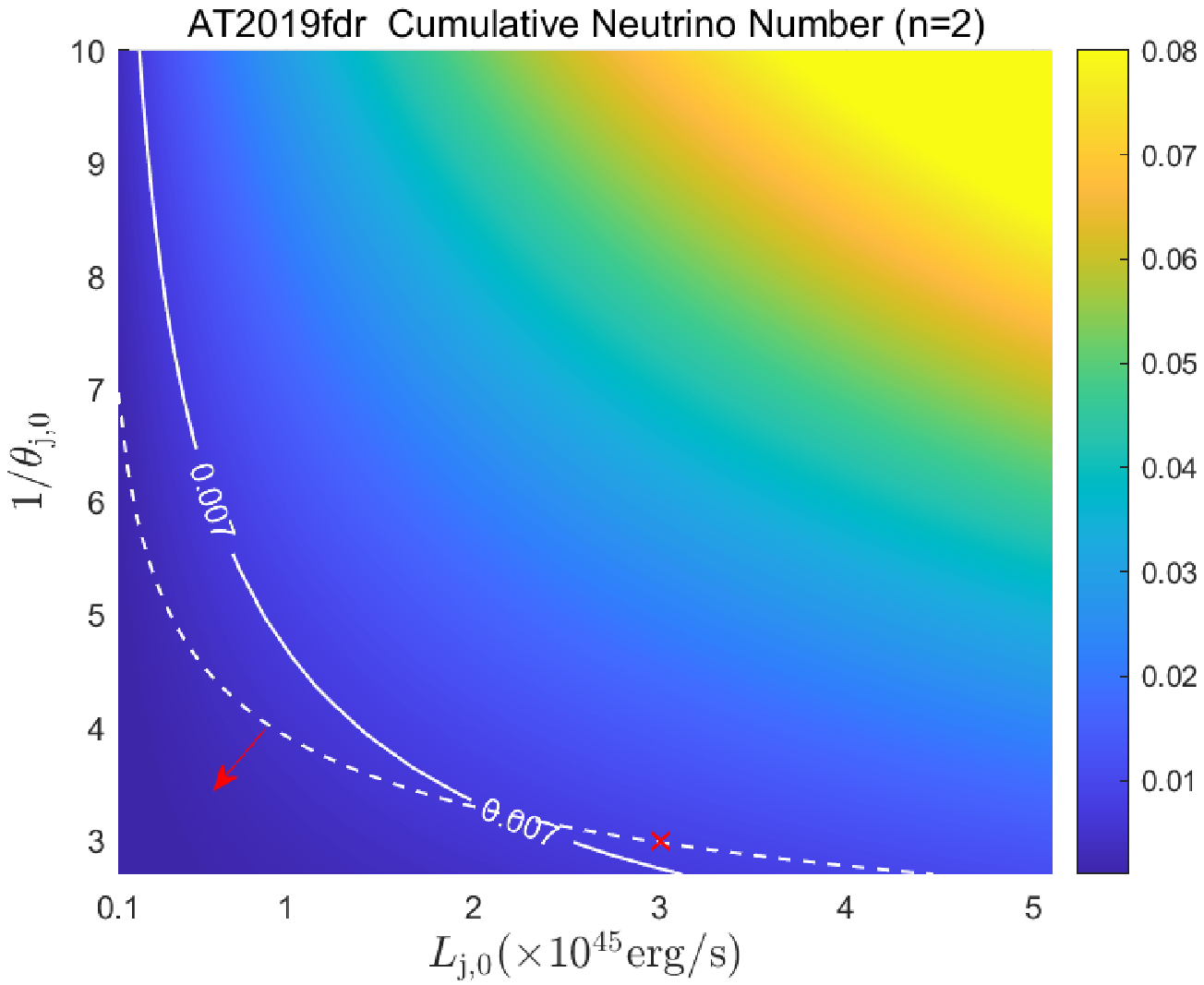}{0.33\textwidth}{(e)}
          \fig{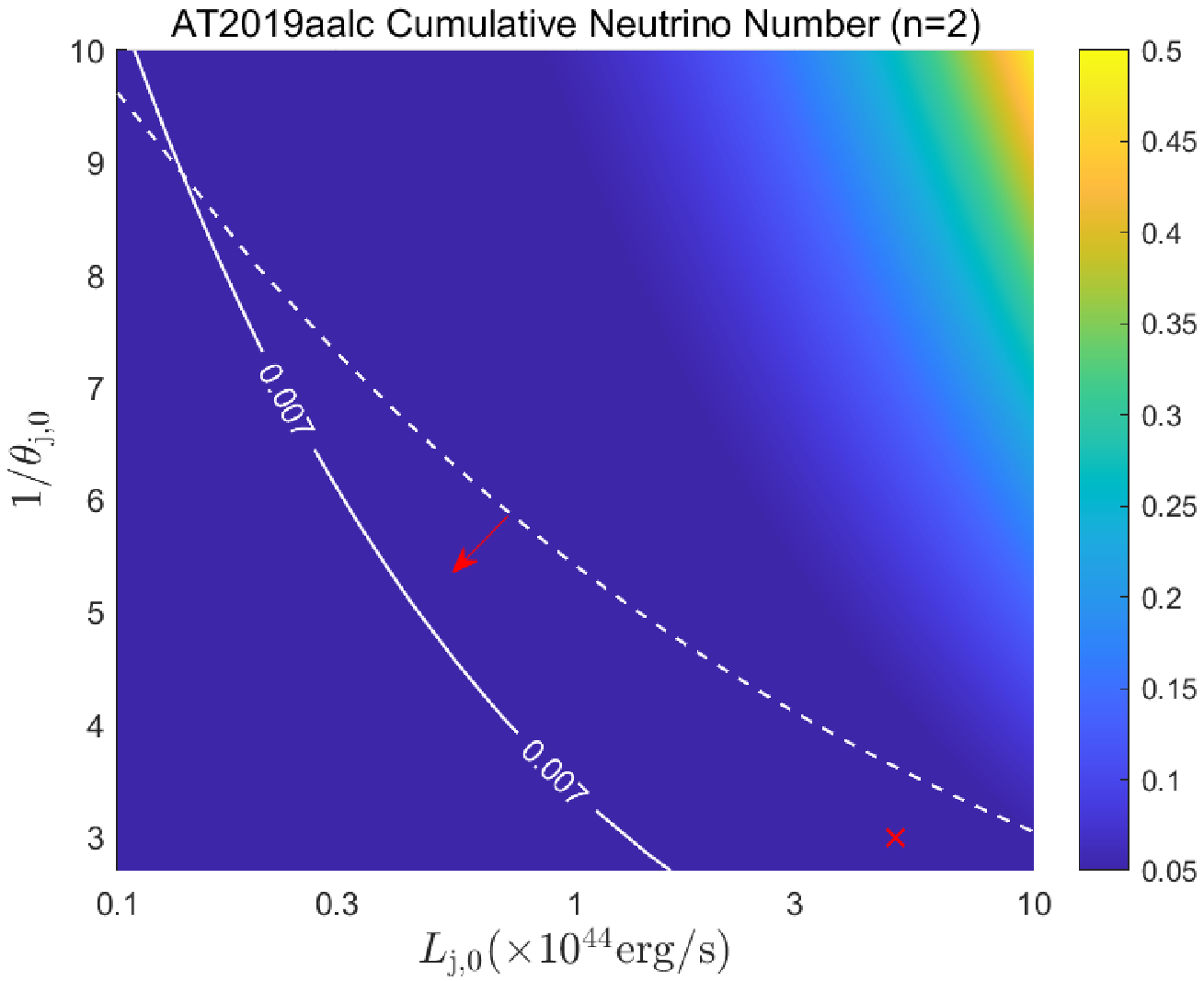}{0.33\textwidth}{(f)}
          }
\caption{ Same as Figure \ref{fig:paratest}, but using the envelope masses of three TDEs given in Case B.
}
\label{fig:paratest2}
\end{figure*}

To calculate the neutrino flux, we need to know the temperature of the thermal photons in the envelope.  The photosphere temperature can be obtained from observations.
To obtain the interior temperature, we need to solve the energy transferring equation, which is
\begin{equation}
\label{rad}
    \frac{du}{dr}=-\frac{3\kappa\rho(r)L_{\rm bol}}{4\pi cr^2},
\end{equation}
where $L_{\rm bol}$ is the bolometric luminosity, $u$ is the energy density and $\kappa=\kappa_{\rm es}=0.33 {\rm cm^2/g}$ is free electron scattering opacity. Adopting the temperature of ideal photon gas $T=(uc/4\sigma)^{1/4}$ where $\sigma$ is Stefan-Boltzmann constant, the solution of the temperature inside the envelope is \citep{Roth2016ApJ...827....3R}
\begin{equation}
\label{temperature}
    T(r,t)={T_{\rm ph}}(t){\left[ {\frac{{3{\tau _{\rm ph}}}}{{4\left( {n + 1} \right)}}\left( {\frac{{R_{\rm ph}^{n + 1}}}{{{r^{n + 1}}}} - 1 + \frac{{4\left( {n + 1} \right)}}{{3{\tau _{\rm ph}}}}} \right)} \right]^{1/4}},
\end{equation}
where $\tau_{\rm ph}\equiv\kappa_{\rm es}\rho_{\rm ph}R_{\rm ph}$. The density at photosphere $\rho_{\rm ph}$ can be obtained by $\int^{R_{\rm ph}}_{R_{\rm T}}\rho(r)dr=M_{\rm env}$, where $R_{\rm T}=R_{\star}(M_{\rm BH}/M_{\star})^{1/3}$ is the disruptive radius. 

Noting that the solution is applicable for any thermal equivalent envelopes since Eq.\ref{rad} is valid in thermal equilibrium. Therefore, Eq.\ref{temperature} can be applied to a time-evolving photosphere. We assume that the envelope temperature decrease with time following the relations  $T_{\rm ph}(t)=T_{\rm ph,peak}(1+t/\tau)^{-\beta}$. The fiducial value $\beta=5/12$ is taken from the simulation \citep{Ryu52020ApJ...904...68K}.
The temperature follows $T\propto r^{-(n+1)/4}$ when $r\ll R_{\rm ph}$. The large number density of thermal photons leads to  $t^{-1}_{p\gamma}\gg t^{-1}_{pp}$. Therefore, the main production channel of neutrinos is $p\gamma$ interaction. The typical energy of protons interacting with thermal photons is $\epsilon_{\rm p}=0.15{\rm GeV}^2/3kT=0.6-6{\rm PeV}$ for $T=10^5-10^6 {\rm K}$ and the neutrino energy is correspondingly $\epsilon_{\nu}=0.05\epsilon_{\rm p}=30-300{\rm TeV}$.

We use a power-law proton spectrum $dN/dE\propto E^{-p}$ with $p=2$. The maximum energy of protons is obtained by equating the acceleration time $t_{\rm acc}=\gamma_{\rm p}m_{\rm p}c/\eta_{\rm acc}eB$ and the cooling time $t_{p\gamma}$, where $\eta_{\rm acc}=0.1$ is the acceleration efficiency, and the magnetic field is given by $B=\sqrt{2\varepsilon_{\rm B}L_{\rm j,iso}/R^2_{\rm int}c}$.  The maximum energy is determined by the balance between the cooling time of $p\gamma$ interactions and the acceleration.  The fluence of neutrinos is
\begin{equation}
    \epsilon_{\nu}\mathcal{F}_{\nu}\approx\int^{t_{\rm d}/(1+z)}_{0}\frac{L_{\rm p}f_{\rm p\gamma}f_{\rm \mu,sup}}{32\pi D^2_{\rm L}\ln(\epsilon_{\rm p,max}/\epsilon_{\rm p,min})}e^{\frac{-\epsilon_{\rm p}}{\epsilon_{\rm p,max}}}dt,
\end{equation}
where $L_{\rm p}=\varepsilon_{\rm p}L_{\rm j,iso}$ is the isotropic proton luminosity and $f_{\rm \mu,sup}=1-\exp{(-t^{-1}_{\rm \mu,dec}/t^{-1}_{\rm \mu,syn})}$ is the suppression factor for muon decay. $t_{\rm \mu,dec}=\gamma_{\mu}\tau_{\mu,\rm dec}$ is the decay time of relativistic muons where $\tau_{\mu,\rm dec}$ is the mean lifetime of muons in the rest frame and $t_{\rm \mu,syn}$ is the synchrotron cooling time of relativistic muons. The minimum energy of protons is $\epsilon_{\rm p,min}=\Gamma m_{\rm p}c^2$. We take the equipartition factors $\varepsilon_{\rm B}=0.1$ and $\varepsilon_{\rm p}=0.2$ in the calculation.

Then we calculate the cumulative muon neutrino number between 10 TeV and 1PeV.
\begin{equation}
    N_{\nu}=\int^{1 {\rm PeV}}_{10 {\rm TeV}}A_{\rm eff}(\epsilon_{\nu})\mathcal{F}_{\nu}d\epsilon_{\nu}.
\end{equation}
The effective area of detectors is $\overline{\nu}_{\mu}$ effective area taken from \cite{Blaufuss2019ICRC...36.1021B}.

\section{Application to AT2019fdr, AT2019dsg and AT2019aalc} \label{sec:app}

To calculate the break-out time of jets and neutrino flux in the three TDEs, we need first to know the envelope mass of each TDE. We suggest two approaches to estimate the envelope mass, one uses the photosphere radius  given by   
$R_{\rm ph}\simeq 10^{15} {\rm cm} \left({M_{\rm env}}/{0.5M_{\odot}}\right)^{1/2}$
for a radiation-pressure supported envelope or a steady state wind outflow \citep{Loeb1997ApJ...489..573L,Roth2016ApJ...827....3R}, and the other uses the bolometric energy $E_{\rm bol}\approx L_{\rm acc,max}\tau\propto M_{\star}$ \citep{Metzger2016MNRAS.461..948M}.
We define the above two cases as Case A and Case B, respectively.

AT2019dsg has a photosphere radius of  $R_{\rm ph}=5.3\times 10^{14}{\rm cm}$, which is common in TDEs, while AT2019fdr and AT2019aalc are rare events with photosphere radii an order of magnitude larger. In case A, the inferred  envelope masses are $0.1M_{\odot}$ for AT2019dsg, $21 M_{\odot}$ for AT2019fdr and $9 M_{\odot}$ for AT2019aalc. The decrease of the event rate with the photosphere radius has been  interpreted  as being due to the lower number of high-mass stars, as implied by the initial mass functions of stars \citep{Velzen2021ApJ9084V}. Here we have assumed that the opacity is the same for three TDEs. In reality, the opacity of AT2019fdr and AT2019aalc could be higher since their photosphere temperatures  are lower\citep{Roth2016ApJ...827....3R},  and then their envelope masses will be correspondingly lower than the above estimates.  

The bolometric  energy of AT2019fdr and AT2019dsg are $3.4\times10^{52}{\rm erg}$ and $1.4\times10^{51}{\rm erg}$, respectively \citep{2019fdr,Velzen2021ApJ9084V}. Since the maximum luminosity and duration of AT2019aalc is about two times higher than that of AT2019dsg, we estimate the bolometric energy of AT2019aalc is around $\sim6\times10^{51}{\rm erg}$. Thus, in case B, the new scaling relation suggests envelope masses of about $0.4 M_{\odot}$, $10 M_{\odot}$ and  $2 M_{\odot}$for AT2019dsg,  AT2019fdr and  AT2019aalc, respectively. 

The black hole mass $M_{\rm BH}$ is estimated from the optical spectrum of the host galaxy or the flare (see Table 1 of \cite{Velzen2021arXiv211109391V}).
We summarize the masses of the disrupted stars and the black holes, as well as other relevant parameters,  for three TDEs in Table \ref{parameters}. The jet lifetime $\tau$  can then be obtained using Eq.3, which is also listed in Table \ref{parameters}.

The cumulative neutrino numbers of three TDEs are shown by the color intensity  in Figures \ref{fig:paratest} and \ref{fig:paratest2}, respectively, for case A and case B.  The solid lines show the parameter values corresponding to the minimum  neutrino number of each TDE required by observations at  $90\%$ confidence\citep{2019dsg2021NatAs...5..510S}. The space  between the solid line and the dashed line (along the direction of the arrow) represents the allowed range  for the jet luminosity and the initial half open angle of the jet, where neutrinos above the minimum  number will be produced while the jets are choked before break-out.

\begin{table}
  \centering
 \caption{Properties and parameters}
 \begin{tabular}{cccc}
 \hline
 \hline
  & AT2019dsg & AT2019fdr & AT2019aalc\\
 \hline

 redshift & 0.051 & 0.267  & 0.0356 \\
 $\epsilon_{\nu}$(TeV) & 217 & 82  & 176 \\
 $t_{\rm d}({\rm days})$\tablenotemark{a}    & 150     &  393         & 148 \\
 $L_{\rm peak}(10^{44}{\rm erg/s})$ & 1.8\tablenotemark{b} & 14\tablenotemark{c}    & 2.7\tablenotemark{d} \\
 $T_{\rm ph,peak}({\rm K})$&  30900\tablenotemark{b}  &    13500\tablenotemark{c}         & 11000\tablenotemark{d}  \\
 $R_{\rm ph}({\rm 10^{14}{\rm cm}})$&  5.3\tablenotemark{b}  &    78         & 51  \\
 $\rm M_{\rm BH}(10^{6}M_{\odot})$& 5\tablenotemark{b}   & 35\tablenotemark{e}     &  20\tablenotemark{e}  \\
 $\tau({\rm days})$    & 40/90 \tablenotemark{b,f}    & 180          & 140 \\
 $\rm M_{\rm env}(M_{\odot})$ in case A& 0.1   & 21      &  9  \\
 $\rm M_{\rm env}(M_{\odot})$ in case B& 0.4   & 10      &  2  \\
 $\rm R_{\rm T}({\rm 10^{13}{\rm cm}})$ in case A& 0.8   & 6.2      &  4.1  \\
 $\rm R_{\rm T}({\rm 10^{13}{\rm cm}})$ in case B& 1.1   & 5.2      &  2.8  \\

 \hline
 \end{tabular}
 \begin{flushleft}
 \tablenotetext{a}{\footnotesize\cite{Velzen2021arXiv211109391V}}
    \tablenotetext{b}{\footnotesize\cite{Velzen2021ApJ9084V}}
    \tablenotetext{c}{\footnotesize\cite{2019fdr}}
    \tablenotetext{d}{\footnotesize\cite{Winter2023ApJ...948...42W}}
    \tablenotetext{e}{\footnotesize \cite{Velzen2021arXiv211109391V}}
    \tablenotetext{f}{\footnotesize  40-days is obtained by fitting observations of bolometric light curves \citep{Velzen2021ApJ9084V}, while 90-days is calculated  from Eq.3. We use $\tau$=40 days in our calculations.}
 \end{flushleft}
 \label{parameters}
\end{table}

AT2019fdr is an exceptionally luminous TDE candidate with  a peak optical/UV bolometric luminosity of $10^{45}{\rm erg/s}$ and a long duration ($\geq1000\,{\rm days}$).   It can be seen from Figure \ref{fig:paratest} that the cumulative neutrino number in the choked jet model can reach $N_{\nu}=0.018$ for a large envelope mass (Case A). With a lower envelope mass (Case B), as shown in Figure \ref{fig:paratest2},  the cumulative neutrino number is at most $N_\nu=0.009$, which falls slightly below the minimum required value. Given that the minimum required value of neutrinos corresponds to a $90\%$ confidence, the choked jet model is still possible for AT2019fdr. 

\begin{figure}
    \centering
    \includegraphics[width=0.50\textwidth]{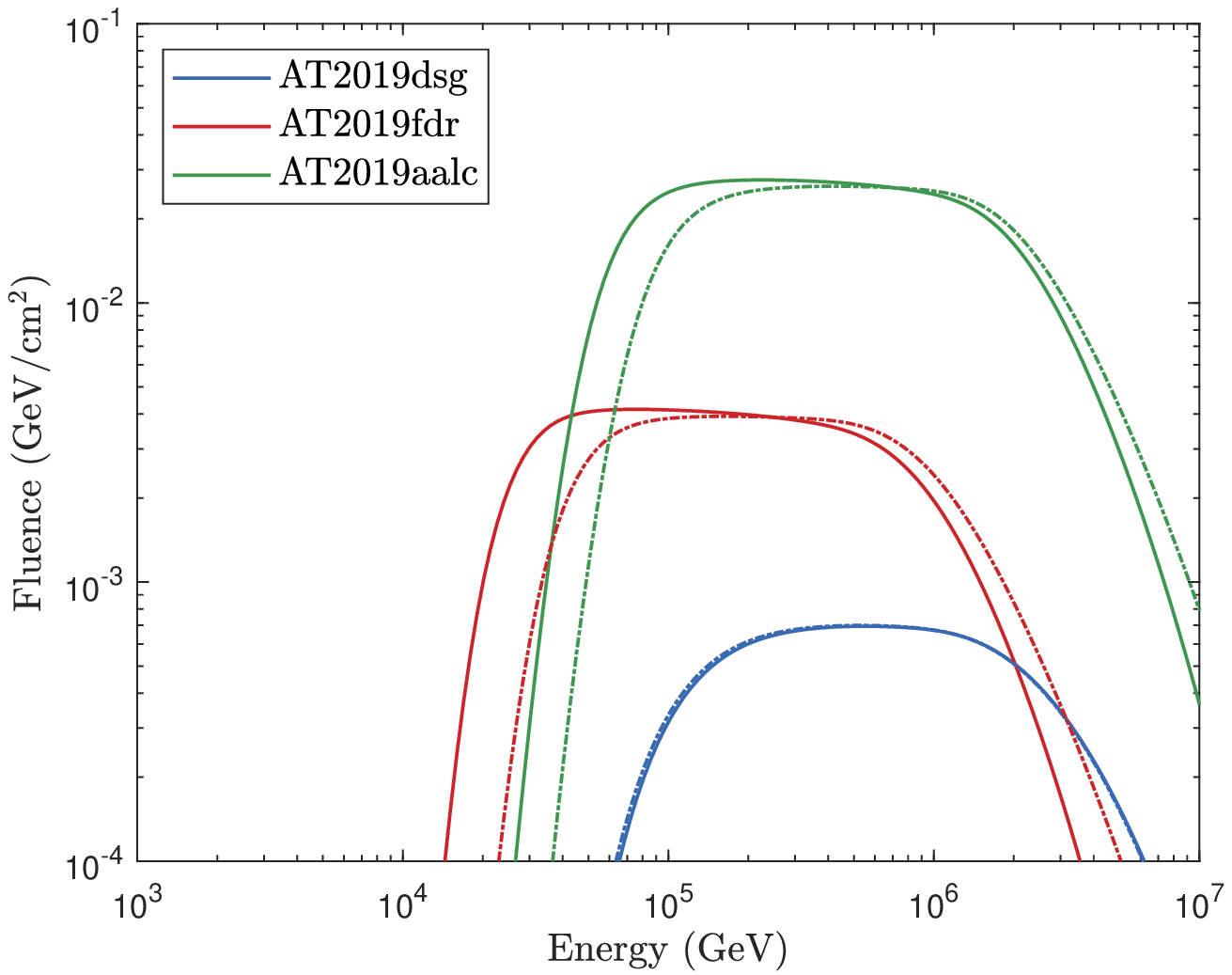}
    \includegraphics[width=0.50\textwidth]{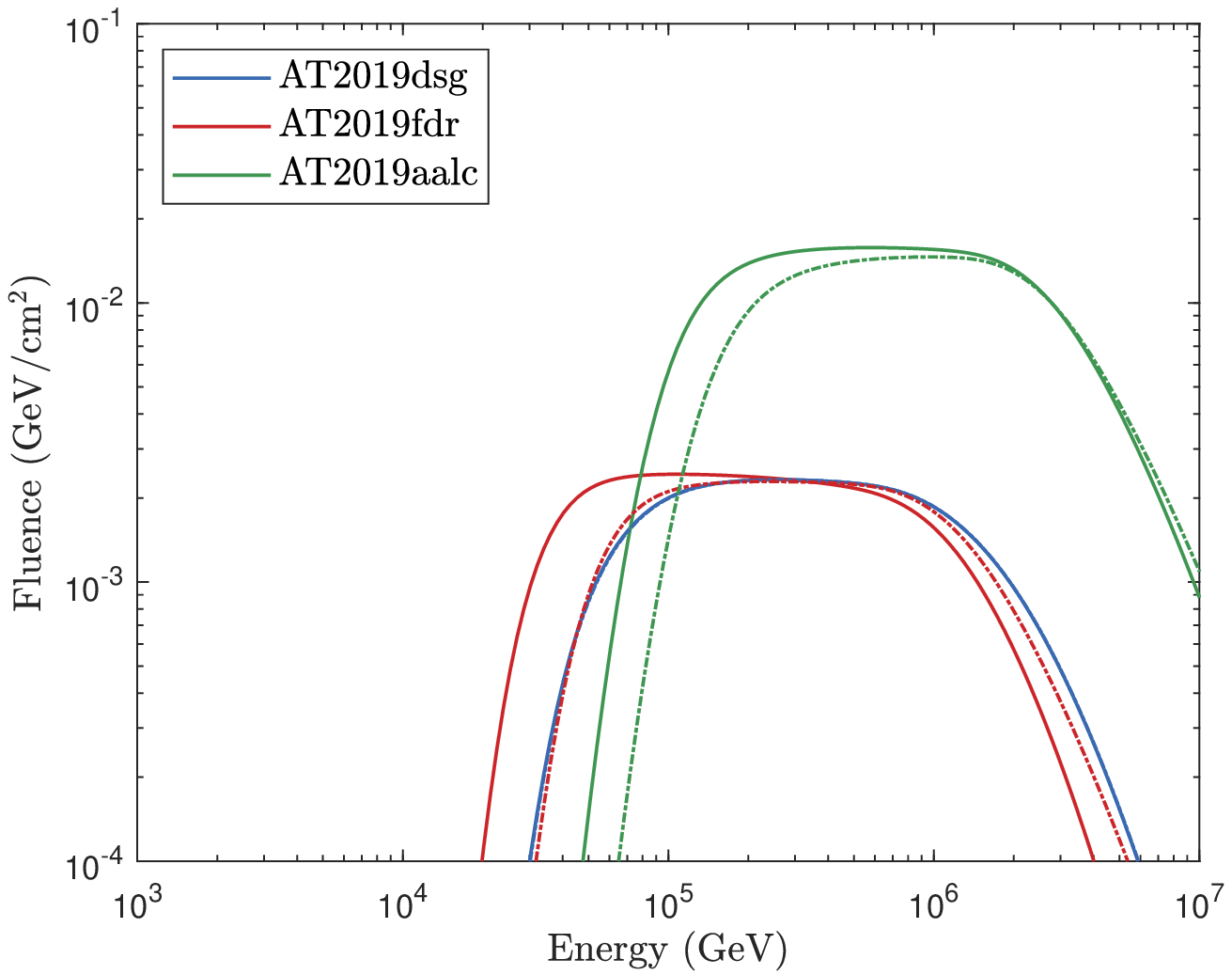}
    \caption{  
    \textit{Upper panel}: Neutrino  spectra of three TDE events in Case A. The jet luminosities are taken to be $L_{\rm j,0}=10^{44}{\rm erg/s}$ for AT2019dsg, $L_{\rm j,0}=5\times10^{45}{\rm erg/s}$ for AT2019fdr and $L_{\rm j,0}=8\times10^{44}{\rm erg/s}$ for AT2019aalc.
    \textit{Lower panel}: Neutrino  spectra of three TDE events in Case  B. The jet luminosities are taken to be $L_{\rm j,0}=3\times10^{44}{\rm erg/s}$ for AT2019dsg, $L_{\rm j,0}=3\times10^{45}{\rm erg/s}$ for AT2019fdr and $L_{\rm j,0}=5\times10^{44}{\rm erg/s}$ for AT2019aalc.
     The solid and dot-dashed lines  represent the neutrino spectra in the $n=3$ and  $n=2$ profiles, respectively.  
}
    \label{fig:flu}
\end{figure}

For AT2019dsg, its small photosphere radius implies a very small mass of the envelope ($0.1M_{\odot}$) in Case A.  The jet can break out easily in this case and only very weak jets will be choked. This leads to a cumulative neutrino number of $N_\nu=0.004$ in the optimistic case, which falls below the expected range ($0.008\la N_\nu\la 0.76$)\citep{2019dsg2021NatAs...5..510S}. Thus, the choked jet model is disfavored for AT2019dsg if the envelope mass is indeed  as small as $0.1M_{\odot}$. In Case B, with an envelope mass of $0.4M_\odot$, the optimal neutrino number of AT2019dsg can reach $N_{\nu}=0.012$ in a small parameter space, satisfying the  requirement of observations \citep{2019dsg2021NatAs...5..510S}.

Similar to AT2019fdr, AT2019aalc is also a long-lasting TDE ($\geq700{\rm days}$) and has a large photosphere radius.  The neutrino number of AT2019aalc can reach $N_{\nu}=0.21$ in Case A and $N_{\nu}=0.08$ in Case B.

\begin{figure}
    \centering
    \includegraphics[width=0.50\textwidth]{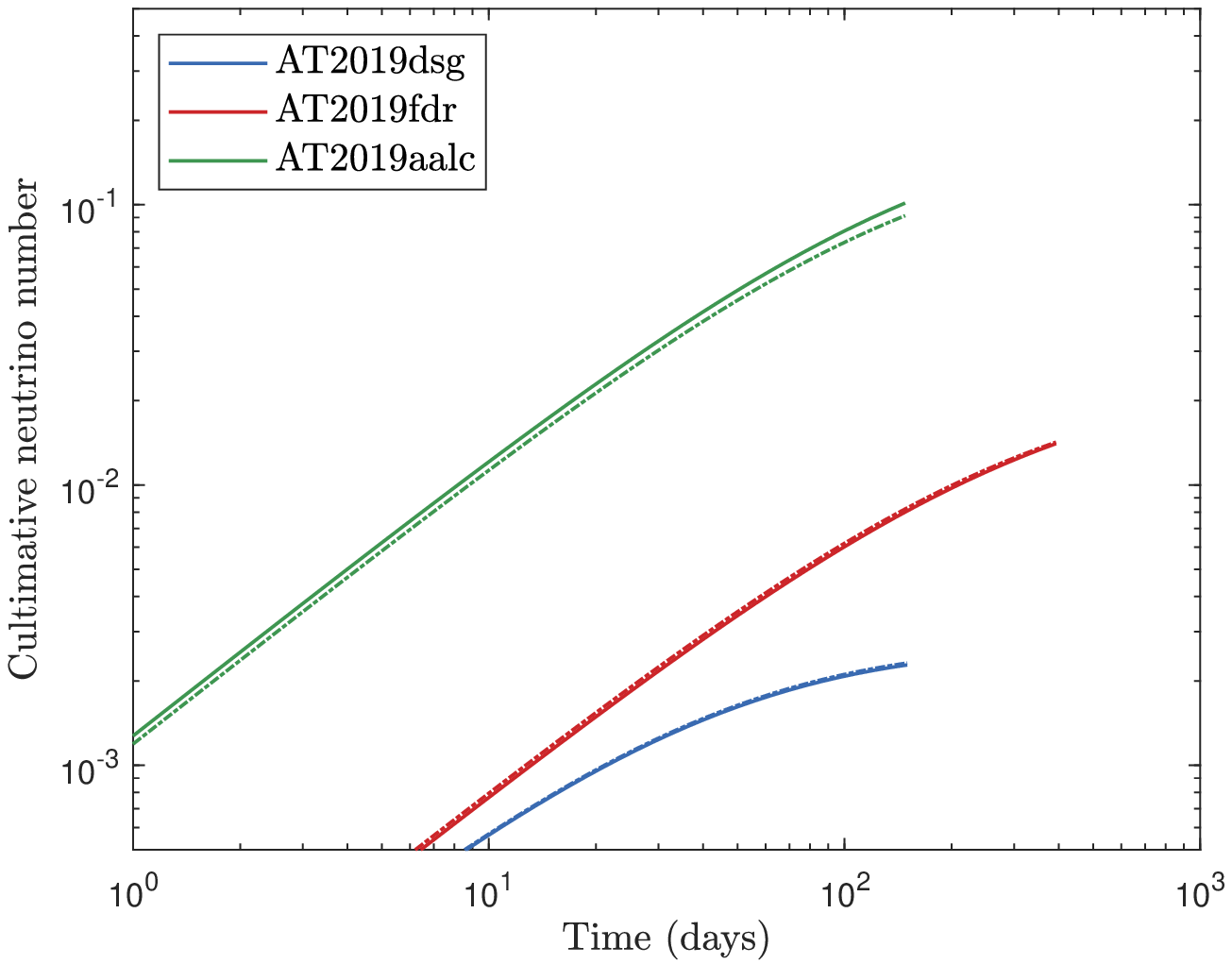}
    \includegraphics[width=0.50\textwidth]{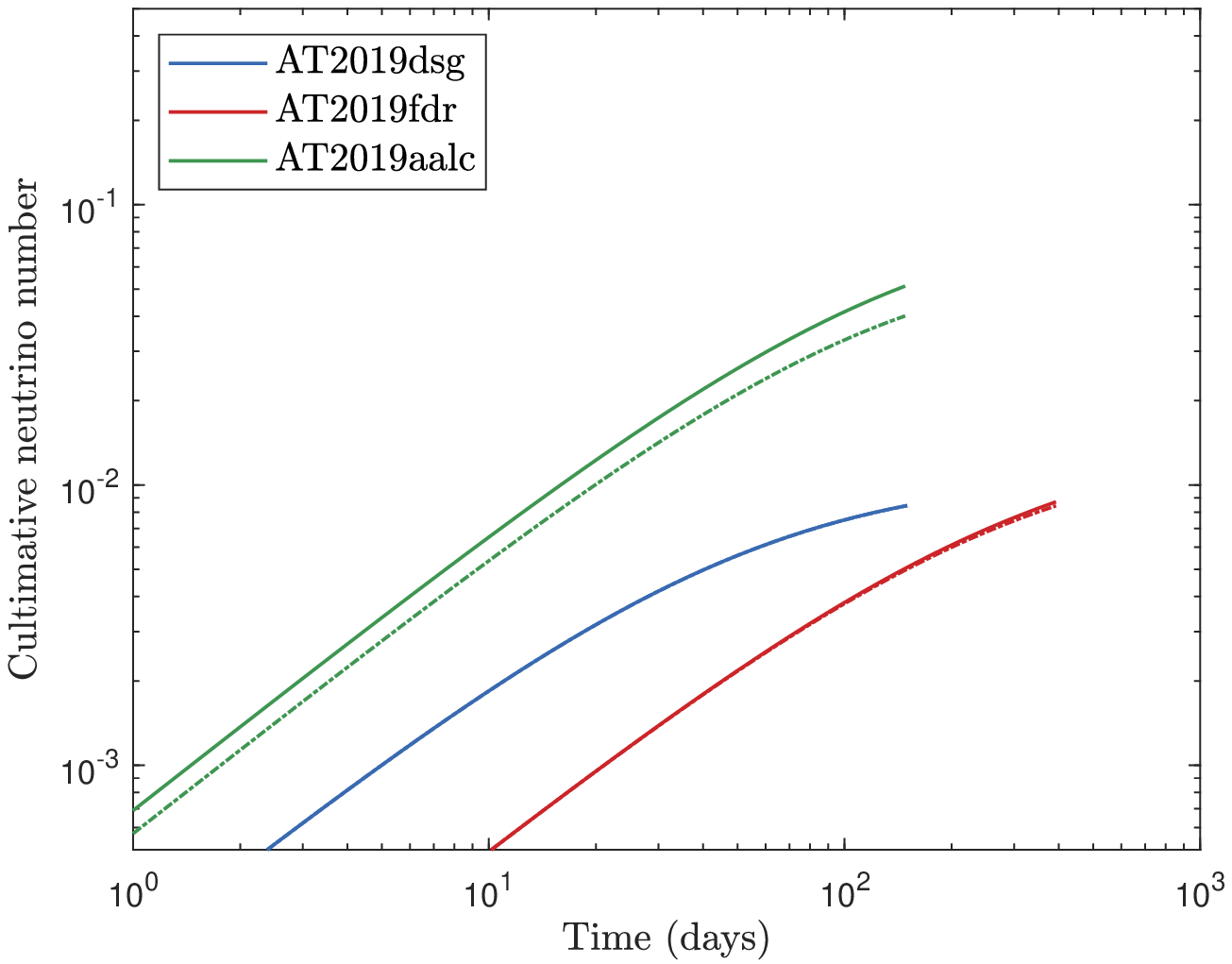}
    \caption{Time evolution of cumulative neutrino numbers. The symbols and parameter values  are identical with those in Fig.\ref{fig:flu}.}
    \label{fig:num}
\end{figure}

The neutrino spectra for three TDEs are shown in Figure \ref{fig:flu} for both Case A (upper panel) and Case B (bottom panel).  The high temperature of $\sim 10^5 {\rm K}$ inside the envelope leads to a neutrino energy of the order of hundreds of TeV, which is  consistent with the observed values of three neutrinos. Because of the high $p\gamma$ efficiency with $f_{p\gamma}\simeq 1$ for neutrino production, the neutrino spectrum follows the proton spectrum and is thus flat above the critical energy. 

The evolution of the  cumulative neutrino numbers with time is shown in  Figure \ref{fig:num}. The cumulative neutrino number increases with time up to a point corresponding to about  $\sim 2\tau$ before flattening. This is because although the jet luminosity starts to decrease after $\tau$, the time-integrated neutrino flux still increases. This can naturally explain the time delay between the neutrino arrival time and the optical peak time of the TDEs.

\section{Conclusions and discussions}
\label{sec:CS}
We have proposed that the neutrino emission associated with three non-jetted TDEs can be explained by the choked jet model, where  relativistic jets are choked inside the quasi-spherical, optically  thick envelope formed from stellar debris of TDEs. From the point of view of observations,  the presence of an envelope can solve the puzzle that the temperatures (few $10^4$ K) found in optically discovered TDEs are significantly lower than the predicted thermal temperature ($> 10^5$ K) of the accretion disk \citep{Gezari2012Natur.485..217G,Arcavi2014ApJ...793...38A}.    While powerful jets, such as that in Swift  J1644+57, can break out from the envelope, less powerful jets would be choked inside the envelope. Cosmic-ray protons accelerated by shocks in choked jets produce high-energy neutrinos via $p\gamma$ interactions with the thermal photons in the  envelope. The  energy of neutrinos produced by choked jets is typically around hundreds of TeV, consistent with the measured energy of three neutrinos. The hundreds of days time delay between neutrino arrival time and TDE optical peaks can be explained as the propagation time of the jets before being choked in the envelope.  The cumulative neutrino numbers  in our model are consistent with the expected range for individual TDEs and can reach $N_{\nu}=0.21$ in  the case of AT2019aalc. One important advantage of the choked jet model is that the neutrino flux is magnified by   the  beaming effect due to the relativistic bulk motion.

The non-detection or weak  radio emission of the three TDEs disfavor the presence of strong relativistic jets \citep{Velzen2021arXiv211109391V,2019fdr}, but are consistent with the choked jet model.  For choked TDE jets, as the neutrino production site is within the optically thick region, the associated high-energy gamma-rays cannot escape. Instead, high-energy gamma-rays are absorbed by low-energy  photons in the envelope, depositing their energy finally into the envelope. Therefore, these choked TDE jets are hidden sources of  gamma-rays, consistent with the non-detection of GeV emission from these TDEs \citep{Velzen2021arXiv211109391V}. The late-time  appearance of thermal X-ray emission in AT2019fdr and AT2019aalc after the neutrino trigger  supports that the accretion discs are obscured by some optically thick materials at earlier time. The X-rays can only leak out after the obscuring gas has  become transparent to X-rays \citep{Metzger2016MNRAS.461..948M,Lu2018ApJ...865..128L}.

For AT2019dsg,  the rapid decrease of the X-ray flux after the first $\sim 40$ days can be explained by an increasing X-ray obscuration \citep{2019dsg2021NatAs...5..510S}, which could be due to the gradual formation of the envelope. The non-detection of any jet emission at early time in AT2019dsg may be due to that the jet formation is also delayed \citep{Cendes2022ApJ...938...28C}. In addition, the neutrino emission from AT2019dsg could be due to successful on-axis or off-axis jets or other processes \citep{Winter2021NatAs...5..472W,Liu2020PhRvD.102h3028L, Murase2020ApJ...902..108M,Wu2022MNRAS.514.4406W}.

Large envelope masses are required for AT2019fdr and AT2019aalc (especially in case A), which would result in  large masses of the disrupted stars. For such large mass stars, the  TDE rate is $\sim 500$ times lower than  the rate for solar-mass-star TDEs (see Figure 3 in \citealt{Kochanek2016bMNRAS.461..371K}). The rate is also dependent on star formation history and the initial mass function. Given a total number of $\sim 100$  TDE candidates have been found so far \citep{TDEreview2021ARA&A..59...21G}, the detection of a few large-mass-star TDEs is not unreasonable, especially considering that luminosities of these TDEs are significantly higher.

\begin{acknowledgments}
The authors thank the referee for the constructive report and thank Ore Gottlieb, Kohta Murase and Tacho Ryu for useful discussions, Zhiyu Zhang for useful comments. This work is supported by the National Key R\&D Program of China under the Grant No. 2018YFA0404203, the National Natural Science Foundation of China (grant numbers 12121003, U2031105), China Manned Spaced Project (CMS-CSST-2021-B11). 
\end{acknowledgments}

\bibliography{reference}{}
\bibliographystyle{aasjournal}



\end{document}